\documentclass[conference]{IEEEtran}
\usepackage[fleqn]{amsmath}
\usepackage{bm}
\usepackage{amssymb,amsfonts}
\usepackage{algorithmic}
\usepackage{graphicx}
\usepackage{textcomp}
\usepackage{xcolor}
\usepackage[utf8]{inputenc}
\usepackage[english]{babel}

\graphicspath{ {./images/} }

\def\BibTeX{{\rm B\kern-.05em{\sc i\kern-.025em b}\kern-.08em
    T\kern-.1667em\lower.7ex\hbox{E}\kern-.125emX}}
\begin{document}

\title{An Upper Bound on the State-Space Complexity of Brandubh}

\author{
    \IEEEauthorblockN{
        Kiernan Compy$^{1}$, Alana Evey$^{1}$, Hunter McCullough$^{1}$, Lindsay Allen$^{1}$, and Aaron S. Crandall$^{2}$
    }

    \IEEEauthorblockA{
        \textit{$^{1}$School of Electrical Engineering and Computer Science, Washington State University, Pullman, WA}\\
        \textit{$^{2}$School of Engineering \& Applied Science,} 
        \textit{Gonzaga University, Spokane, WA}\\
        Corresponding Email: hunter.mccullough@wsu.edu, crandall@gonzaga.edu
%        Email: hunter.mccullough@wsu.edu, alana.evey@wsu.edu, kiernan.compy@wsu.edu, lindsay.allen@wsu.edu, and crandall@gonzaga.edu
    }
}

\IEEEoverridecommandlockouts\IEEEpubid{\begin{minipage}{\textwidth}\ \\[12pt]
978-1-7281-4533-4/20/\$31.00 \copyright 2020 IEEE
\end{minipage}}

\maketitle
\begin{abstract}
Before chess came to Northern Europe there was Tafl, a family of asymmetric strategy board games associated strongly with the Vikings. The purpose of this paper is to study the combinatorial state-space complexity of an Irish variation of Tafl called Brandubh. Brandubh was chosen because of its asymmetric goals for the two players, but also its overall complexity well below that of chess, which should make it tractable for strong solving. Brandubh's rules and characteristics are used to gain an understanding of the overall state-space complexity of the game. State-spaces will consider valid piece positions, a generalized rule set, and accepted final state conditions. From these states the upper bound for the complexity of strongly solving Brandubh is derived. Great effort has been placed on thoroughly accounting for all potential states and excluding invalid ones for the game. Overall, the upper bound complexity for solving the game is around $10^{14}$ states, between that of connect four and draughts (checkers).

\end{abstract}


\section{Introduction}
Tafl refers to a family of asymmetric games played on an $n\times n$ square checkered board, similar to chess~\cite{Harding2010}. These Tafl games were most popular until about the 12$^{th}$ century when they were largely supplanted by chess~\cite{Murray1952}, though they were observed being played in Lapland as late as the 1700's~\cite{Linnaeus1811}. The rules to Tafl games were only recorded in a few places, and are continually undergoing revision for both their historical state and to build a balanced rule set for today's players~\cite{Ashton,Walker2014,Davidson2011}.

\begin{figure}[tbp]
 \centerline{\includegraphics[width=.9\columnwidth]{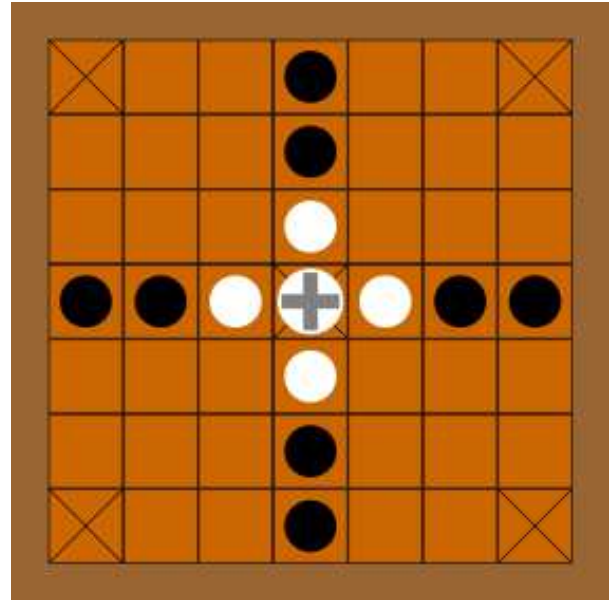}}
 \caption{Brandubh board initial piece layout. Black are attackers, white are defenders, with the King in the center.}
 \label{fig:tafl_board}
\end{figure}

What makes Tafl games asymmetric and different from chess is the ratio of pieces on each side. The attackers outnumber the defenders by $2:1$. The defenders begin in the middle of the board surrounded by attackers, the organization of which differs in the Tafl variants. The goal of the defenders is to break out of the attacker's encirclement as to have the king reach one of the four corners, or to eliminate all attackers, while conversely the attackers are trying to complete the encirclement to capture the king. These games are similar to Fox Games~\cite{Pentagram1990}. For the purpose of this paper, the $7\times7$ Brandubh variation shown in Figure~\ref{fig:tafl_board}~\cite{Dantas2007} of Tafl will be considered~\cite{Walker2014}.

The Tafl family of  games is not well studied from a theoretic or complexity perspective, and therefore state complexities are relatively unknown.
These games are asymmetric, though they should be analyzable using symmetric decomposition~\cite{Tuyls2018} to determine advantageous game play for each player.
Given the symmetric nature of the piece movement rules, role-based symmetrization and training agents should be able to perform both roles of attacker and defender~\cite{Cressman2014,Cressman2003}. The symmetric nature of the moves should reduce training time in computer agents using these methods. This becomes especially important as work on Tafl games moves to evaluating proposed rule changes to make the games more balanced by allowing agents to be trained quickly and efficiently as rules change.

Brandubh was chosen from among the other Tafl variants for state-space analysis as it has the simplest rule set, and is played on the smallest board. The simplicity will allow for a more accurate calculation of the upper bound as well as the ability to expand the understanding of other variations. The total complexity will also be most likely to allow for future work to weakly or strongly solving Brandubh.

\section{Motivation}
Our motivation for strongly solving Brandubh is to build and test AI agents of varying difficulties. We hope to eventually extend the agents to include other variations of Tafl board games as well.

The importance of exploring asymmetrical analysis is apparent when considering its widespread applications. Not only is it relevant in developing game theory, but it is also proving useful for medical diagnosis based on neuro-imaging techniques such as MRI and CT scan results~\cite{Kalavathi}.

The work on Tafl games could be valuable when addressing asymmetric game theory analysis.
The most notable concern is in relation to cyber security. The U.S. Military has been creating AIs for weapon systems using asymmetry, which has further tempted the development of adversarial AIs related to it. These are intended to influence false predictions in their machine learning models that will result in unexpected, and possibly lethal behavior~\cite{feldman}. Exploring not only the total complexity of Tafl games, but continued mechanisms of how to build functional AI agents to compete in adversarial asymmetric situations can contribute to security applications.

\section{Related Works}

This work focuses on estimating the total number of game states possible in a game of Brandubh.
The most famous similar work was by Shannon~\cite{shannon1950} in estimating the number of games available in chess.
In Shannon's work, he looked at the number of possible moves, the length of average games, and any rules that would limit the state-space and came to an initial number of $10^{120}$ games.
That number has been reduced through eliminating invalid moves, position transposition, handling ties, and focusing on sensible moves. This has brought the state-space to around $10^{40}$.
Other games have similar analyses performed, such as Connect 4 with about $2^{13}$ positions~\cite{Edelkamp2017}, domineering with $2^{15}$~\cite{Allis} positions, and draughts (checkers) at $10^{20}$ positions~\cite{Schaeffer2007}.
This paper presents a similar analysis for Brandubh, and by extension, the other Tafl games.

There are few papers addressing the state-space analysis of Tafl, or Brandubh specifically.
One such analysis was done by Slater, who roughly estimates Hnefatafl (played on an $11\times11$ board) at $10^{43}$~\cite{Slater2015}.
Slater's work did not account for move transpositions and corner states included in this Brandubh analysis.

A more complete complexity analysis of a similar game was done by Galassi, which addressed Tabult~\cite{Galassi2021}. Tabult is a Tafl game played on a board larger than Brandubh, with a $9\times9$ tile board. Galassi divided the move space into subspaces and handled more of the possible state space reductions than Slater's work. Galassi's conclusion is that Tabult's complexity is roughly the same as Draughts (checkers). The authors of this work on Brandubh used some of Galassi's approaches for handling state space reductions in Brandubh.

\begin{figure}[tbp]
  \centerline{\includegraphics{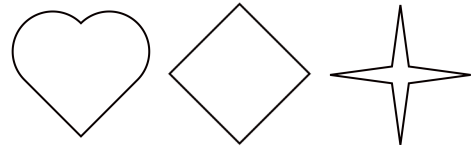}}
  \caption{Game piece representations, in order from left to right: King, defender, attacker.}
  \label{fig:piece_shapes}
\end{figure}

\section{Brandubh Rules Used}
\subsection{Setup}
\begin{figure}[tbp]
  \centerline{\includegraphics[width=0.75\columnwidth]{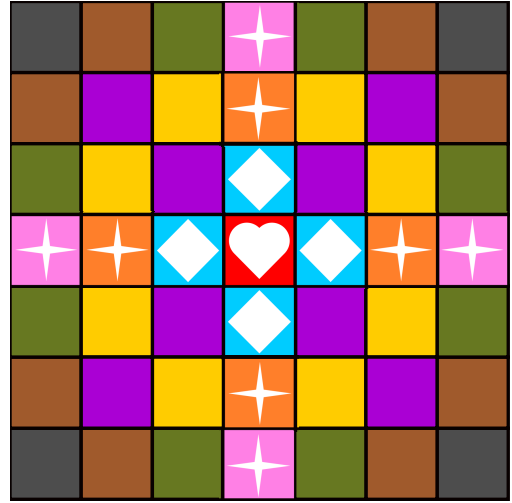}}
  \caption{Brandubh Initial Game Board Layout}
  \label{fig:gameboard}
\end{figure}

For all Tafl variations, the game board is a perfect square. The Brandubh game board, shown in Figure~\ref{fig:gameboard}, can be thought of as a two dimensional array consisting of 7 rows and 7 columns. It has a standardized setup that includes a king, four defenders, and eight attackers. The king always starts in the very center of the board at position $4\times4$. This square of the game board will henceforth be referred to as the throne. The four defenders start the game adjacent to the king, and the 8 attackers fill the remaining spaces in row 4 and column 4. The king will be represented by a heart, the defenders by a diamond, and the attackers by a four point star. Each representation is pictured in Figure~\ref{fig:piece_shapes}.

\begin{figure}[tbp]
  \centerline{\includegraphics[width=0.75\columnwidth]{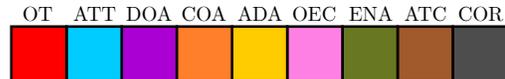}}
  \caption{Color Key for Game Board States}
  \label{fig:gameboard_key}
\end{figure}

\subsection{Board Layout}
All the sub-cases for Brandubh can be described with the colored tiles used in Figure~\ref{fig:gameboard_key}. Each colored tile represents events in which the king is standing on that tile type. OT (Red Tile) is abbreviated for On Throne; situations in which the king is on the throne. ATT (Teal Tiles) is abbreviated for Adjacent To Throne; situations in which the king is adjacent to the throne. OA is abbreviated for Open Area; situations in which the king is not adjacent to an edge, or horizontally/vertically adjacent to the center. This section is divided up into 3 separate sub-cases: DOA (Purple Tiles) is abbreviated for Diagonal Open Area, COA (Orange Tiles) is abbreviated for Center Open Area, and ADA (Yellow Tiles) is abbreviated for Adjacent to Diagonal Open Area. OEC (Pink Tiles) is abbreviate for On Edge Center; situations in which the king is in the center of the edge of the board. ENA (Green Tiles) is abbreviated for Edge Non Adjacent; situations in which the king is on the edge of the board, not a center edge piece, and not adjacent to the corner tile. ATC (Brown Tiles) is abbreviated for Adjacent To Corner; situation in which the king is adjacent to a corner tile. COR (Black Tiles) is abbreviated for Corner; situations in which the king is in the corner tile.

\subsection{Movement}
In Brandubh, movement is standardized. Every piece may move any number of squares vertically or horizontally across the board. Piece jumping is not allowed, and two pieces may not occupy the same space. The king may not return to the throne after leaving it, and both attackers and defenders are never allowed to rest on the throne. The rules of Brandubh state any piece may move through the throne at any time. However, to aid in the simplicity of calculations this was considered an illegal move.

\subsection{Capture}
 A capture is defined as any move which results in the removal of a piece from the board. Across all Tafl games, captures primarily involve trapping a piece between two hostiles, though there are game specific variations. Pieces may only be captured by hostiles which are adjacent to them, and diagonal captures are never allowed. In Brandubh, the edges of the board may not be considered hostile in captures. It should also be noted the side which is making the capture must be the side to initiate the move. The Brandubh variation allows for the king to participate in captures, slightly skewing the $2:1$ ratio. Brandubh also allows for the king to be captured like any other piece, provided it is not resting on the throne or an adjacent square. If the king is on or adjacent to the throne, the king is only captured if surrounded by hostile squares. Each valid capture scenario has been considered and is explained in the following sections.
 
\subsubsection{On Throne} When the king is on the throne, four attacker pieces are required for capture.

\begin{figure}[htbp]
  \centerline{\includegraphics[width=0.25\columnwidth]{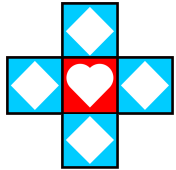}}
  \caption{On Throne Capture}
  \label{fig:KOnThrone}
\end{figure}

\subsubsection{Adjacent to Throne} When the king is in a cell adjacent to the throne, three opposing pieces are required for capture.

\begin{figure}[htbp]
  \centerline{\includegraphics[width=0.25\columnwidth]{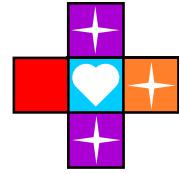}}
  \caption{Adjacent to Throne King Capture}
  \label{fig:KAdjThrone}
\end{figure}

\subsubsection{Hostile Throne} When the throne is not occupied by the King, the throne is considered hostile, and therefore can be used to help capture pieces by either attackers or defenders. 

\begin{figure}[htbp]
  \centerline{\includegraphics[width=0.25\columnwidth]{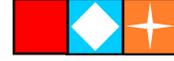}}
  \caption{Hostile Throne}
  \label{fig:ThroneCapt}
\end{figure}

\subsubsection{Off Throne, Off Edge} When any piece is not adjacent to or on the throne and it is not on an edge or corner, it is in an \textbf{open area}. In order for a capture to happen, the side trying to capture must move into a position such that the piece being captured is between two capturing pieces occupying the same row or the same column.

\begin{figure}[htbp]
  \centerline{\includegraphics[width=0.25\columnwidth]{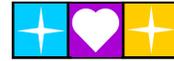}}
  \caption{Open Area Capture}
  \label{fig:Sandwhich}
\end{figure}

\subsubsection{Hostile Corners} A corner piece may be considered hostile by both attackers and defenders.

\begin{figure}[htbp]
  \centerline{\includegraphics[width=0.25\columnwidth]{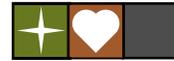}}
  \caption{Hostile Corner Capture}
  \label{fig:HostileCorner}
\end{figure}

\subsection{Win Scenarios}
In Brandubh, and all Tafl games, there are three win conditions:

\begin{enumerate}
    \item Attackers win if the king is captured
    \item Defenders win if the king reaches a corner square
    \item Defenders win if all attackers are captured
\end{enumerate} 

Additionally, if the same board state is repeated three times in a row, the game is a draw.

\section{Calculations}
\subsection{Na{\"i}ve State-space Complexity}

A na{\"i}ve approach to calculating the state-space complexity of Brandubh would be to consider every possible state stemming from the initial state. We call the rough estimation \bm{$UB_{naive}$}, which is described as:
\begin{align}\label{ub_naive}
    UB_{naive} &=2^5 \times 4^{44}\\\nonumber&\approx 9.91\times10^{27}
\end{align}

The purpose of the calculations within this paper are to calculate the most accurate upper bound possible by considering as few illegal moves as possible. The simplifications made to $UB_{naive}$ are explained in the proceeding sections.

\subsection{Rotations and Mirroring}

The game board used by Brandubh is a perfect square, and as such can be divided into four equally sized quadrants. This equality can be used to simplify $UB_{naive}$ into what we will refer to as \bm{$UB_{tight}$}, by treating mirrors or rotations of each quadrant as one state. That is, if a piece occupies the square $3\times2$; through mirroring it is the same as occupying square $3\times6$ or $5\times2$, or through rotation it is the same as occupying square $5\times6$. 

\subsection{Definitions}

 In order to simplify the calculation of states, the board has been broken down into sections. The sections to be considered are the throne, the corners, the cell the king occupies, and the cells adjacent to the king. The various types of game pieces will be referenced by the following variables:
 \begin{itemize}
     \item Attackers: \boldsymbol{$a$}
     \item Defenders: \boldsymbol{$d$}
 \end{itemize}

The variable \boldsymbol{$k$} will represent all the tiles on the board minus the four corners, the throne, the tile the king is on, and the king's adjacent tiles. The following equation will be used to calculate the set of possible state variations, given the variables $a$, $d$, and $k$.
 
\begin{align}
        P_{k}^{(a,d)} = \frac{k!}{a!d!(k-a-d)!}
\end{align}

We will use a single variable Kronecker delta, \boldsymbol{$\delta_{n}$}, to denote if there is at least one attacker adjacent to the King. The necessity of $\delta_{n}$ becomes clear when accounting for capture states. The function is as follows:
\begin{align}
    \delta_{n}=\delta_{0,n}=
    \begin{cases}
        1 &\mbox{if } n = 0 \\
        0 &\mbox{if } n \neq 0
    \end{cases}
\end{align}

% NOTE: 0 to n is open spaces or number of pieces?

Next, let \boldsymbol{$a_r$} be the number of hostiles adjacent to the king and \boldsymbol{$d_r$} the number of defenders. Note that $a_r$ and $d_r$ can individually range from 0 to $n$, where \boldsymbol{$n$} is the total number of open spaces adjacent to the king. Similarly $a_r+d_r$ can range from 0 to $n$. 
Now, consider the case where the king is surrounded, $a_r = n$. Let \boldsymbol{$A$} be a single row vector where $A_n$ represents the number of unique arrangements that make this possible. A single row vector is needed for simplifying equation (\ref{eq:BK/WK}). With $a_r$ surrounding attackers and $d_r$ total defenders. Let \boldsymbol{$D$} be a 2 dimensional vector where $D_{a_r,d_r}$ represents the number of unique arrangements of the surrounding attackers and defenders.
Now, given $n$ spaces around the king and $k$ spaces non-adjacent to the king that are occupiable by either attackers or defenders, let $f(n, k, A, D)$ calculate the number of unique states that can arise:

\begin{align}
    f(n,k,A,D)=\sum_{a_r=0}^{n-1}A_{a_r}\sum_{d_r=0}^{n-a_r}D_{a_r,d_r}\sum_{a=\delta_{a_r}}^{8-a_r}\sum_{d=0}^{4-d_r}P_{k}^{(a,d)}
    \label{eq:BK/WK}
\end{align}

\subsection{Upper Bound of Non-End Sates}\label{test_label}

% NOTE: I changed defenders to attackers here "and each column the number of attackers in the same range"

$D$ is populated by the values of $a_r$ and $d_r$. Each column of the matrix represents the number of defenders being considered from 0 to 4, and each column the number of attackers in the same range. To make the process used for all non-end state calculations clear, the process for the first case is explicitly shown. However, the function specified in (\ref{eq:BK/WK}) will be used to represent the remainder of the cases.

\begin{itemize}
    \item \textbf{On Throne (OT)} \\
    OT is the case where the king has not left its initial position. The throne is shown as the red tile in Figure~\ref{fig:gameboard}. The calculations for all states given this assertion are as follows:
    \begin{align}
        D=
        \begin{bmatrix}
        1 & 1 & 2 & 1 & 1 \\
        1 & 2 & 2 & 1     \\
        1 & 2 & 1         \\
        1 & 1             \\
        0                 \\
        \end{bmatrix}
        A=\left[1,1,2,1,0\right]
    \end{align}
     
     Let \boldsymbol{$i$} be the count of attackers adjacent to the king. Due to the rule of capturing a king in the center tile requiring 4 attackers surrounding him, $i$ has a maximum value of 3. \boldsymbol{$C_{i}$} will denote the possible cases where the king is surrounded. In all sub-cases where the king is on the throne and not captured, $n=4$ and $k=40$. The components of on throne are:
    
    \begin{align}\label{on_throne_0}
        C_{0}&=D_{0,0}\sum_{a=1}^8 \sum_{d=0}^{4} P_{40}^{(a,d)}+D_{0,1}\sum_{a=1}^8 \sum_{d=0}^{3} P_{40}^{(a,d)} \\
        &+D_{0,2}\sum_{a=1}^8 \sum_{d=0}^{2} P_{40}^{(a,d)}+D_{0,3}\sum_{a=1}^8 \sum_{d=0}^{1} P_{40}^{(a,d)} \nonumber \\
        &+D_{0,4}\sum_{a=1}^8 P_{40}^{(a,d)}\nonumber\\\nonumber
        &\approx4.99 \times 10^{12}
    \end{align}
    
    \begin{align}\label{on_throne_1}
        C_{1}&=D_{1,0}\sum_{a=0}^7 \sum_{d=0}^{4} P_{40}^{(a,d)}+D_{1,1}\sum_{a=0}^7 \sum_{d=0}^{3} P_{40}^{(a,d)} \\
        &+D_{1,2}\sum_{a=0}^7 \sum_{d=0}^{2} P_{40}^{(a,d)}+D_{1,3}\sum_{a=0}^7 \sum_{d=0}^{1} P_{40}^{(a,d)} \nonumber\\
        &\approx1.44 \times 10^{12}
    \end{align}
    
    \begin{align}\label{on_throne_2}
        C_{2}&=D_{2,0}\sum_{a=0}^6 \sum_{d=0}^{4} P_{40}^{(a,d)}+D_{2,1}\sum_{a=0}^6 \sum_{d=0}^{3} P_{40}^{(a,d)} \\
        &+D_{2,2}\sum_{a=0}^6 \sum_{d=0}^{2} P_{40}^{(a,d)} \nonumber\\
        &\approx 3.14\times10^{11}
    \end{align}
    
    \begin{align}\label{on_throne_3}
        C_{3}&=D_{3,0}\sum_{a=0}^5 \sum_{d=0}^{4} P_{40}^{(a,d)}+D_{3,1}\sum_{a=0}^5 \sum_{d=0}^{3} P_{40}^{(a,d)}\\
        &\approx 5.16\times10^{10}
    \end{align}
    \begin{align}
        OT&=A_0C_{0}+A_1C_{1}+A_2C_{2}+A_3C_{3}\\\nonumber
        &\approx 7.11\times10^{12}
    \end{align}
    \item \textbf{Adjacent To Throne (ATT)}
    \newline ATT refers to all states where the king is adjacent to the throne, and is not being captured. These positions are represented as blue tiles in Figure~\ref{fig:gameboard}. The calculations for all possible states for ATT are as follows:
    \begin{align}
        D=
        \begin{bmatrix}
        1 & 3 & 3 & 1 \\
        1 & 2 & 1     \\
        1 & 1         \\
        1             \\
        \end{bmatrix}
        A=\left[1,2,2,1\right]
    \end{align}
    \begin{align}
    ATT&=f(4,40,A,D)\\\nonumber
    &\approx9.63\times10^{12}
    \end{align}
    
    \item \textbf{Open Area (OA)}
    \newline OA refers to all states where the king is not along the edge, adjacent to the corner, adjacent to the throne, and is not being captured. The calculations for OA have been broken up into 3 unique parts; \textbf{Diagonal Open Area (DOA), Adjacent to Diagonal Open Area (ADA)} and \textbf{Center Open Area (COA)}. The breakdown into three sections reduces redundancy by taking advantage of the symmetry of the game board, and hence makes the overall upper bound calculation more accurate. The three state space calculations will be summed together and referred to as OA for the remainder of the paper. As seen below each of these sub-cases share the same D matrix with different A vectors. 
    \begin{align}
        \begin{split}
            D=
            \begin{bmatrix}
            1 & 4 & 6 & 4 & 1 \\
            1 & 3 & 3 & 1     \\
            1 & 2 & 1         \\
            1 & 1             \\
            0
            \end{bmatrix}
        \end{split}
        \begin{split}
            A_{DOA}&=\left[1,2,4,2,0\right]\nonumber\\
        A_{ADA}&=\left[1,4,6,4,0\right]\nonumber\\
        A_{COA}&=\left[1,3,4,3,0\right]\nonumber
        \end{split}
    \end{align}
    
    DOA  is represented by the purple tiles in Figure~\ref{fig:gameboard} and can be calculated with the following summations:
    \begin{align}
        DOA &= 2 \times f(4,39,A_{DOA},D)\\\nonumber
        &\approx 1.68 \times 10^{13}
    \end{align}
    
    ADA is represented by the yellow tiles in Figure~\ref{fig:gameboard} and can be calculated with the following summations:
    \begin{align}
        ADA &= f(4,39,A_{ADA},D)\\\nonumber
        &\approx 1.14 \times 10^{13}
    \end{align}
    
    COA is represented by the orange tiles in Figure~\ref{fig:gameboard} and can be calculated with the following summations:
    \begin{align}
        COA &= f(4,39,A_{COA},D)\\\nonumber
        &\approx 9.63 \times 10^{12}
    \end{align}
    
    Using the above calculations gives:
    \begin{align}
        OA&=DOA + ADA + COA\\\nonumber
        &\approx3.78\times10^{13}
    \end{align}
    
   \item \textbf{On Edge Not Adjacent to Corner (OE)}
   \newline OE defines the state where the king is along the edge of the board, but not adjacent to the corners or in a captured state. This section is broken up into 2 subsections to take advantage of mirroring along the center-line. These will be called \textbf{On Edge Not Adjacent to Corner Center (OEC)} and \textbf{On Edge Not Adjacent to Corner Not Center (ENA)}. Each of these sub-cases share the same matrix $D$ with differing $A$ vectors:
    \begin{align}
        \begin{split}
            D=
            \begin{bmatrix}
            1 & 3 & 3 & 1 \\
            1 & 2 & 1     \\
            1 & 1         \\
            1
            \end{bmatrix}
        \end{split}
        \begin{split}
            A_{OEC}=\left[1,2,2,1\right]\nonumber\\
            A_{ENA}=\left[1,3,3,1\right]\nonumber
        \end{split}
    \end{align}
    
    OEC is represented by the pink tiles in Figure~\ref{fig:gameboard}. The calculations are:
    \begin{align}
        OEC &= f(3,40,A_{OEC},D) \\\nonumber
        &\approx 9.63\times10^{13}
    \end{align}
    
    ENA is represented by the army green tiles in Figure~\ref{fig:gameboard} and can be calculated by:
    \begin{align}
        ENA &= f(3,40,A_{ENA},D)\\\nonumber
        &\approx 1.14\times10^{13}
    \end{align}
    
    Using the calculations above gives:
    \begin{align}
        OE&=OEC+ENA\\\nonumber
        &\approx2.10\times10^{13}
    \end{align}

    \item \textbf{Adjacent To Corner (ATC)}
    \newline ATC refers to when the king is adjacent to a corner, but not in a captive state. ATC is shown as a brown tile in Figure~\ref{fig:gameboard}, and its calculation is as follows:

    \begin{align*}
        D=
        \begin{bmatrix}
        1 & 2 & 1 \\
        1 & 1     \\
        1         \\
        \end{bmatrix}
        A=\left[1,2,1\right]
    \end{align*}
    \begin{align}
    ATC&=f(3,41,A,D)\\\nonumber
    &\approx1.14\times10^{13}
    \end{align}
\end{itemize}

\subsection{Upper Bound of End States}

In addition to all possible positions of the king where the game does not end, those which lead to a win condition must be considered. Unlike in the non-end state calculations, matrices do not simplify calculations for end state conditions, and therefore will not be used or shown. End state conditions are more complex than non-end state conditions and as such require more than one function to represent them. The following combination will be used to reduce the redundancy which would otherwise be included in to the Upper Bound End States. The variable $d_r$ is still used to represent the number of defenders adjacent to the king, and $u$ represents how many defenders \emph{could} be adjacent to the king, with a maximum value of 4.

\begin{align}
    \binom{u}{d_r}=\frac{u!}{d_r!(u-d_r)!}
\end{align}

The variables \bm{$y$} and \bm{$t$} will be used to represent the maximum number of attacker and defender pieces which are not adjacent to the king. 

\begin{equation}
        g(y,t,k)=\sum_{a=0}^y \sum_{d=0}^{t} P_{k}^{(a,d)}
\end{equation}

Variable \bm{$j$} represents possible transformations:

\begin{equation}
    x(j,t,k)=j\left[ \sum_{d=0}^{t} P_{k}^{(0,d)}+\sum_{d=0}^{t-1} P_{k}^{(0,d)}+k\sum_{d=0}^{t-1} P_{k-1}^{(0,d)}\right]
\end{equation}

\begin{align}
    h(q,k) &= \bigg[\sum_{d_r=0}^1 \binom{1}{d_r} \sum_{a=\delta_{q}}^{8-q} \sum_{d=0}^{4-d_r} P_{k}^{(a,d)}\nonumber\\
    &+\sum_{a=0}^{8-q-1} \sum_{d=0}^{4} P_{k}^{(a,d)}\bigg]
\end{align}

\begin{itemize}
    \item \textbf{King Surrounded On Throne (KOT)}
    \newline This case covers all possible combinations of valid capture states while the King is on the throne.
    \begin{align}
        KOT&=g(4,4,40)\nonumber\\&\approx6.91\times10^{9}
    \end{align}
    
    \item \textbf{King Surrounded Adjacent To Throne (KAT)}
    \newline This case covers all cases where the king is captured while adjacent to the throne.
    \begin{align}
        KAT&=4\times g(5,4,40)\nonumber\\&\approx4.61\times10^{10}
    \end{align}
    
    \item \textbf{No Attackers Left (NAL)}
    \newline NAL is the win scenario where all attackers have been captured. For simplification, this condition has been split into the possible scenarios leading to this state.
    \newline
    \newline
    Final attacker being captured with one defender or the king, and a hostile square (\textbf{CE}):
    \begin{align}
        CE&=x(2,4,42)\\\nonumber
        &=1241244
    \end{align}
   Final attacker being captured with two defenders or a defender and the king (\textbf{CNE}):
    \begin{align}
        CNE&=x(11,4,41)\\\nonumber
        &=506495
    \end{align}
    
    Using the calculations above gives:
    \begin{align}
        NAL&=CE+CNE\\\nonumber
        &=1747739
    \end{align}
    \item \textbf{King Surrounded by Two Pieces (KS)}
    \newline 
    KS refers to the scenario where the king is captured by two attackers on either of its sides as depicted in Figure~\ref{fig:Sandwhich}. This case is completely different from the others within this section, as it takes place in the open area, and therefore has a unique calculation as shown below:
    \begin{align}
        KS &= 4 \bigg[\sum_{d_r=0}^2 \binom{2}{d_r} \sum_{a=0}^6 \sum_{d=0}^{4-d_r} P_{39}^{(a,d)}\nonumber\\
        &+\sum_{d_r=0}^1 \binom{1}{d_r} \sum_{a=0}^5 \sum_{d=0}^{4-d_r} P_{39}^{(a,d)}\nonumber\\
        &+\sum_{a=0}^4 \sum_{d=0}^{4} P_{39}^{(a,d)}\bigg]\\\nonumber
        &\approx2.59\times10^{12}
    \end{align}
    
    \item \textbf{King Captured On Edge (KCE)}
    \newline KCE finds all situations in which the king is captured on the edge of the board, but not in the corner or adjacent to corner.
    \begin{align}
        KCE&=2\times h(2,40)\nonumber\\&\approx6.53\times10^{11}
    \end{align}
    
    \item \textbf{King Adjacent to Corner (KAC)}
    \newline KAC captures every instance of the king being captured while adjacent to the corner tile, as depicted in Figure~\ref{fig:HostileCorner}.
    \begin{align}
        KAC&=h(1,41)\nonumber\\&\approx2.03\times10^{12}
    \end{align}
    
    \item \textbf{King Corner (KC)}
    \newline KC is the win condition in which the king enters any 4 of the board corners. This is represented as a black tile and the following math:
    \begin{align}
        KC&=h(0,42)\nonumber\\&\approx1.14\times10^{13}
    \end{align}
\end{itemize}

\subsection{Report Findings}

\boldsymbol{$UB_{NE}$} is the summation of all potential game states which do not result in the end of the game. The summation is as follows:
\begin{align}
    UB_{NE} &= OT + ATT + OA + OE + ATC\nonumber\\
    &\approx8.68\times10^{13}
\end{align}
\boldsymbol{$UB_{E}$} is the summation all possible end states, and is as follows:
\begin{align}
        UB_{E} &= KOT + KAT + CE \nonumber\\
        &+ CNE + KS + KCE \nonumber\\
        &+ KAC + KC \\\nonumber
        &\approx 1.67\times10^{13}
\end{align}

\boldsymbol{$UB_{tight}$} will be calculated by the addition of all possible states the board can exist in.
\begin{align}
        UB_{tight}&=UB_{NE} + UB_{E}\\\nonumber
        &\approx 1.04\times10^{14}
\end{align}
$UB_{naive}$ was found to be approximately $9.91\times10^{27}$ as shown in equation \eqref{ub_naive}. Our calculations for $UB_{tight}$ came to be approximately $1.04\times10^{14}$. The vast difference between $UB_{naive}$ and $UB_{tight}$ highlights the effects of redundancy on the number of states. 

\section{Conclusion}

The summations presented within this paper represent a careful estimation of upper bound state-space calculations for the Brandubh variant among the family of Tafl games. A combinatorial upper bound using a na{\"i}ve approach of $9.91\times 10^{27}$ was initially found. After applying state-space reductions when accounting for mirrored states, board edges, special king states, and edge cases by applying the game's rules, a tighter bound of approximately $1.04\times 10^{14}$ was achieved.

From the results shown, it can be determined the complexity is roughly between that of connect four~\cite{Japp} and 8x8 domineering~\cite{Allis}. As both connect four and domineering are considered solved games, the conclusion can be drawn that Brandubh is tractable for both weak and strong solving. Checkers was solved for $10^{20}$ positions using hundreds of computers over two decades. Newer computers, faster storage, and being six orders of magnitude smaller than checkers should allow Brandubh to be solvable on a much quicker timeline.

\section{Future Work}

This initial state-space evaluation should allow for updates and variants of the game to be evaluated more quickly. All Tafl games are being changed and updated to find game balance, so this state-space is expected to change in the future.

These techniques and approaches to address various game state cases should be applicable to other Tafl games. Similar games such as Hnefatafl, Tablut, Ard Ri, Tawlbwrdd, and Alea Evangelii are all open to continued analysis, which should follow a similar process to the Brandubh analysis presented here.

Based on this work showing Brandubh's tractability, the authors are working towards a weak and strong solving of the game to establish the rules' gameplay balance. This work should also provide resources to improve AI agent development for the Tafl family of games.

\section{Version Amendment}
In a previous version of this work published in the 2020 IEEE Conference on Games (CoG)~\cite{Compy2020}, the related works citation of Andrea Galassi's work done on Tablut was accidentally left out during the final stages of the drafting process. The authors apologize to Galassi for this oversight.

\bibliographystyle{ieeetr}

\bibliography{mybibliography}

\begin{thebibliography}{10}

\bibitem{Harding2010}
T.~Harding, ``'{A} fenian pastime'? {E}arly {I}rish board games and their
  identification with chess,'' {\em Irish Historical Studies}, vol.~37,
  no.~145, pp.~1--22, 2010.

\bibitem{Murray1952}
H.~J.~R. Murray, {\em A history of board-games other than chess}.
\newblock Clarendon press, 1952.

\bibitem{Linnaeus1811}
C.~Linnaeus, {\em Lachesis lapponica, or a tour in Lapland}.
\newblock White and Cochrane, 1811.

\bibitem{Ashton}
J.~C. Ashton, ``Linnaeus’s game of tablut and its relationship to the ancient
  viking game hnefatafl,'' {\em The Heroic Age: A Journal of Early Medieval
  Northwestern Europe}, vol.~13, pp.~1526--1867, 2010.

\bibitem{Walker2014}
D.~Walker, {\em Reconstructing Hnefatafl}.
\newblock Lulu Com, 2014.

\bibitem{Davidson2011}
N.~Davidson, ``Hnefatafl: an experimental reconstruction.'' Online, 2011.
\newblock Accessed: 2020.05.24.

\bibitem{Pentagram1990}
Pentagram, {\em Pentagames}.
\newblock Fireside, Simon \& Schuster Inc, 1990.

\bibitem{Dantas2007}
L.~Dantas, ``Brandubh board.'' Online, Jan. 2007.
\newblock Released to public domain.

\bibitem{Tuyls2018}
K.~Tuyls, J.~Pérolat, M.~Lanctot, G.~Ostrovski, R.~Savani, J.~Z. Leibo,
  T.~Ord, T.~Graepel, and S.~Legg, ``Symmetric decomposition of asymmetric
  games,'' {\em Scientific Reports}, vol.~8, p.~1015, Jan. 2018.

\bibitem{Cressman2014}
R.~Cressman and Y.~Tao, ``The replicator equation and other game dynamics,''
  {\em Proceedings of the National Academy of Sciences}, vol.~111,
  no.~Supplement 3, pp.~10810--10817, 2014.

\bibitem{Cressman2003}
R.~Cressman, C.~Ansell, and K.~Binmore, {\em Evolutionary dynamics and
  extensive form games}, vol.~5.
\newblock MIT Press, 2003.

\bibitem{Kalavathi}
P.~Kalavathi, M.~Senthamilselvi, and V.~B.~S. Prasath, ``Review of
  computational methods on brain symmetric and asymmetric analysis from
  neuroimaging techniques,'' {\em Technologies}, vol.~5, no.~2, p.~16, 2017.

\bibitem{feldman}
P.~Feldman, A.~Dant, and A.~Massey, ``Integrating artificial intelligence into
  weapon systems,'' 2019.

\bibitem{shannon1950}
C.~E. Shannon, ``Programming a computer for playing chess,'' {\em The London,
  Edinburgh, and Dublin Philosophical Magazine and Journal of Science},
  vol.~41, no.~314, pp.~256--275, 1950.

\bibitem{Edelkamp2017}
S.~Edelkamp, ``Bdds for minimal perfect hashing: Merging two state-space
  compression techniques.'' Online, 2017.

\bibitem{Allis}
L.~V. Allis {\em et~al.}, {\em Searching for solutions in games and artificial
  intelligence}.
\newblock Ponsen \& Looijen Wageningen, 1994.

\bibitem{Schaeffer2007}
J.~Schaeffer, N.~Burch, Y.~Bj{\"o}rnsson, A.~Kishimoto, M.~M{\"u}ller, R.~Lake,
  P.~Lu, and S.~Sutphen, ``Checkers is solved,'' {\em Science}, vol.~317,
  no.~5844, pp.~1518--1522, 2007.

\bibitem{Slater2015}
J.~Slater, ``On tafl: state space complexity.'' Online, 2015.
\newblock Accessed: 2020.05.24.

\bibitem{Galassi2021}
A.~Galassi, ``An upper bound on the complexity of tablut,'' 2021.

\bibitem{Japp}
H.~[van~den Herik], J.~W. Uiterwijk, and J.~[van Rijswijck], ``Games solved:
  Now and in the future,'' {\em Artificial Intelligence}, vol.~134, no.~1,
  pp.~277--311, 2002.

\bibitem{Compy2020}
K.~Compy, A.~Evey, H.~McCullough, L.~Allen, and A.~S. Crandall, ``An upper
  bound on the state-space complexity of brandubh,'' in {\em 2020 IEEE
  Conference on Games (CoG)}, pp.~519--525, IEEE, 2020.

\end{thebibliography}
 %\nocite{*}  % This includes all uncited references in the .bib file. Useful in some places, but not here.

\end{document}